# Digital Filters for Instantaneous Frequency Estimation

**Hugh L. Kennedy**

Hugh.Kennedy@DEWC.com; Hugh.Kennedy@UniSA.edu.au; Hugh.L.Kennedy@gmail.com

## Abstract

This technical note is on digital filters for the high-fidelity estimation of a sinusoidal signal's frequency in the presence of additive noise. The complex noise is assumed to be white (i.e. uncorrelated) however it need not be Gaussian. The complex signal is assumed to be of (approximately) constant magnitude and (approximately) polynomial phase such as the 'chirps' emitted by bats, whale 'songs', pulse-compression radars, and frequency-modulated (FM) radios, over sufficiently short timescales. Such digital signals may be found at the end of a sequence of analogue heterodyning (i.e. mixing and low-pass filtering), down to a bandwidth that is matched to an analogue-to-digital converter (ADC), followed by digital heterodyning and sample rate reduction (optional) to match the clock frequency of the processor. The spacing of the discrete frequency bins (in cycles per sample) produced by the Fast Fourier Transform (FFT) is equal to the reciprocal of the window length (in samples). However, a long FFT (for fine frequency resolution) has a high complexity and a long latency, which may be prohibitive in embedded closed-loop systems, and unnecessary when the channel only contains a single sinusoid. In such cases, and for signals of constant frequency, the conventional approach involves the (weighted) average of instantaneous phase differences. General, naïve, optimal, and pragmatic (recursive), filtering solutions are discussed and analysed here using Monte-Carlo (MC) simulations.

# Contents

# Executive Summary

This technical note is on digital filters for the high-fidelity estimation of a sinusoidal signal's frequency in the presence of additive noise. The complex noise is assumed to be white (i.e. uncorrelated) however it need not be Gaussian. The complex signal is assumed to be of (approximately) constant magnitude and (approximately) polynomial phase such as the 'chirps' emitted by bats, whale 'songs', pulse-compression radars, and frequency-modulated (FM) radios, over sufficiently short timescales.

Such digital signals may be found at the end of a sequence of analogue heterodyning (i.e. mixing and low-pass filtering), down to a bandwidth that is matched to an analogue-to-digital converter (ADC), followed by digital heterodyning and sample rate reduction (optional) to match the clock frequency of the processor. The spacing of the discrete frequency bins (in cycles per sample) produced by the Fast Fourier Transform (FFT) is equal to the reciprocal of the window length (in samples). However, a long FFT (for fine frequency resolution) has a high complexity and a long latency, which may be prohibitive in embedded closed-loop systems, and unnecessary when the channel only contains a single sinusoid. In such cases, and for signals of constant frequency, the conventional approach involves the (weighted) average of instantaneous phase differences. For this approach, Kay (1989) derived an optimal weight that minimises the error variance of the frequency estimator, for a finite window of a given length. This weight is designed to whiten the colour that is introduced to the noise by the two-point differencing operator involved in the transformation from complex to angle (i.e. instantaneous phase-difference) domains. The operator is a first-order high-pass moving-average system, realised through the application of the arc tangent operator to the product of the complex waveform and its complex conjugate that is delayed by one sample. Thus, the optimal phase-weight is gradually tapered towards the edges of the finite window, for a (low-pass) frequency response with lower sidelobes where the gain of the phase differencing operator is elevated.

Averaging over long time-intervals is necessary in low signal-to-noise ratio (SNR) environments and the many (real) arithmetic operations may be prohibitive at very high data-rates; therefore, cheaper recursive alternatives are discussed and analysed here, using Monte-Carlo (MC) simulation at high and low SNR, in ideal and non-ideal conditions. Cascaded Integrator Comb (CIC) filters, and Cascaded Leaky Integrator (CLI) filters, are considered.

The (low pass) frequency responses of the various averaging filters are used to account for the errors observed in the MC simulations. It is shown that the white-noise gain (WNG) of the high-pass and low-pass filters in series (a band-pass system) is equal to the error variance of the frequency estimator (for the assumed ideal signal). The WNG is equal to the sum of the squared impulse response, or the integral of the squared magnitude response, of the digital band-pass filter.

The finite impulse-response (FIR) of the symmetric two-sided taper applied by the CIC filter is generated by convolving multiple rectangular weights, realized by placing $K_{\text{CIC}}$ integrator-comb components in series, with the marginally stable pole of the integrator (on the unit circle at $z = 1$) cancelled by a zero of the comb. This two-sided taper is symmetric, for perfect phase linearity; however, it has relatively high sidelobes which amplify high-frequency coloured-noise and interference.

The infinite impulse response (IIR) of the CLI filter is similarly realized by placing $K_{\text{CLI}}$ (first order) 'leaky' integrator components in series, each with an exponentially decaying impulse response and a stable (real) pole (inside the unit circle at $z = p$). The asymmetric impulse response $h[m]$, of this filter is the discrete Erlang function $c_w m^\kappa p^m$, where $c_w$ is a normalizing factor. It rises as a monomial from $m = 0$: i.e. as $m^\kappa$, where $\kappa$ is the shape parameter, with $\kappa = K_{\text{CLI}} - 1$; and decays as an exponential towards $m \to \infty$: i.e. as $p^m = e^{-m/\lambda}$, where lambda is the scale parameter and $p = e^{-1/\lambda}$; for a frequency response $H(e^{i\omega})$, with greater high-frequency attenuation at the expense of imperfect pass-band phase-linearity.

Both CIC and CLI filters are crude approximations of the Kay weight, which is optimal for the somewhat abstract problem defined above (a single sinusoid in white noise). However, such conditions are rarely encountered in practice. For instance: internal or external (deliberate or accidental) interference, coloured complex noise, phase and magnitude modulation, and non-polynomial phase progression, are almost always present in real environments and systems. And in these situations, phase-averaging filters with lower high-frequency gain and/or wider bandwidth may be required. The two parameters of the Erlang weight provide a simple way to fine tune the response of this low-complexity frequency estimator.

Kay also considered an alternative approach to the estimation problem, whereby the averaging filter is applied *before* the angle operator, i.e. the complex-to-angle transformation, and showed that the error variance of this estimator is almost as low as the standard approach in high-SNR scenarios (for the same filter coefficients); however, it is much worse at low SNR. Both were compared in the MC simulations conducted here and the error of the alternative approach was found to be significantly lower for a sinusoid with randomly generated band-limited phase-and-magnitude modulation. A third approach is also proposed here, whereby the phase differences are dynamically weighted according to their relative magnitudes when they are averaged. The observed error variance for this method in the MC simulations is only slightly worse than Kay's alternative approach.

## List of mathematical symbols

| | | |
|---|---|---|
| $F_s$ | Sampling rate (samples per second or Hertz) | |
| $T_s$ | Sampling period (seconds), $T_s = 1/F_s$ | |
| $\omega$ | (Relative angular) frequency (radians per sample) | |
| $f$ | (Relative normalized) frequency (cycles per sample), $f = \omega/2\pi$ | |
| $\Omega$ | (Angular) frequency (radians per second), $\Omega = F_s \omega$ | |
| $F$ | Frequency (cycles per second or Hz), $F = F_s f$ | |
| $i$ | Imaginary unit | |
| $e$ | Euler's number | |
| $n$ | Sample index | |
| $m$ | Delay index | |
| $l$ | Lag index (an integer delay) | |
| $N$ | Number of samples | |
| $M$ | Number of delays | |
| $L$ | Number of lags | |
| $x$ | The sampled (i.e. measured) waveform (complex) to be processed, signal plus noise | |
| $\psi$ | Complex signal of unknown instantaneous phase (thus frequency, which is to be estimated), $\psi[n] = Ae^{i\theta[n]}$ | |
| $A$ | Magnitude of the complex signal (a real constant) | |
| $\varepsilon_\psi$ | Uncorrelated complex noise (i.e. white, but not necessarily Gaussian), added to the complex signal, i.e. $x = \psi + \varepsilon_\psi$ | |
| $\sigma_\varepsilon^2$ | Variance of the white noise $\varepsilon_\psi$ | |
| $\theta$ | Instantaneous phase of the signal (radians) | |

| | |
|---|---|
| $\tilde{\theta}$ | Raw (i.e. noisy) phase 'measurement' (radians), derived from the application of the angle operator (i.e. the argument operator) to the complex waveform |
| $\bar{\theta}$ | Unwrapped phase-difference measurement |
| $\varepsilon_\theta$ | Phase noise (radians), $\tilde{\theta} = \theta + \varepsilon_\theta$ |
| $K_\theta$ | Order of the instantaneous phase model |
| $\theta_0$ | Phase offset (radians), $\theta_0 = \phi_\psi$ |
| $\theta_1$ | Phase velocity (radians per sample), $\theta_1 = \omega_\psi$ |
| $\theta_2$ | Phase acceleration (radians per sample per sample) |
| $\theta_3$ | Phase jerk (radians per sample per sample per sample) |
| $\theta_k$ | $k$th derivative of the instantaneous phase, with respect to time, for $k = 0 \ldots K_\theta$ |
| $\omega_\psi$ | Frequency of the signal (radians per sample) |
| $f_\psi$ | Frequency of the signal (cycles per sample) |
| $F_\psi$ | Frequency of the signal (radians per second) |
| $\phi_\psi$ | Phase offset of the signal (radians) |
| $\tilde{\omega}$ | Raw (i.e. noisy) phase-difference measurement (radians per sample) |
| $\bar{\omega}$ | Unwrapped phase-difference measurement (radians per sample) |
| $\hat{\omega}_\psi$ | Estimated signal frequency (radians per sample) |
| $\tilde{f}$ | Phase-difference measurement (cycles per sample) |
| $\bar{f}$ | Unwrapped phase-difference measurement (cycles per sample) |
| $\hat{f}_\psi$ | Estimated signal frequency (cycles per sample) |
| $\varepsilon_\omega$ | Phase-difference noise (radians), $\tilde{\omega} = \omega_\psi + \varepsilon_\omega$ |
| $w$ | Weight used to average (thus smooth, via a low-pass filter) the phase differences |
| $y$ | Output of the low-pass filter that is used to compute the average phase difference |
| $x[n]x^*[n-1]$ | Delayed conjugate product |
| $q$ | Nominal pass-band group-delay of the low-pass filter (samples) |
| $P_{\varepsilon_\psi}(\omega)$ | Power spectral density of the complex noise |
| $P_{\varepsilon_\theta}(\omega)$ | Power spectral density of the phase noise |
| $P_{\varepsilon_\omega}(\omega)$ | Power spectral density of the phase-difference noise |
| $\sigma_{\hat{\omega}}^2$ | Variance of the frequency estimate error |
| $\sigma_{\tilde{\omega}}^2$ | Variance of the phase-difference noise |
| $\upsilon$ | White-noise gain of a digital filter |
| $\upsilon_{\text{HPF}}$ | White-noise gain of the high-pass filter, used to generate phase differences |
| $\upsilon_{\text{LPF}}$ | White-noise gain of the low-pass filter, used to average phase differences |
| $\upsilon_{\text{BPF}}$ | White-noise gain of the band-pass system, formed from the high-pass and low-pass filters in series |
| $h[m]$ | Impulse response of a digital filter |
| $H(z)$ | Discrete-time transfer-function of a digital filter, $H(z) = B(z)/A(z)$ |
| $z$ | Coordinate in the complex $z$-plane, reached via the $\mathcal{Z}$-transform |
| $H(e^{i\omega})$ | Frequency response of a digital filter |
| $\boldsymbol{w}$ | Weight vector |
| $\boldsymbol{X}$ | Regressor matrix |
| $\boldsymbol{P}_{\varepsilon_\omega}$ | Covariance matrix of the coloured phase-difference noise |
| $K_{\text{CIC}}$ | Number of cascaded integrator-comb filter components in series |
| $M_{\text{CIC}}$ | Length of the rectangular impulse-response of each integrator-comb component |
| $\kappa$ | Shape parameter of the Erlang weight, an integer that is greater than or equal to zero |
| $\lambda$ | Timescale parameter of the Erlang weight (samples) |
| $c_w$ | Normalizing factor for the Erlang weight |
| $p$ | Smoothing parameter of the Erlang weight; all poles of the Erlang filter are at $z = p$; $p$ is real and on the [0,1] interval (i.e. inside the unit circle) |
| $\mu_w$ | Mean of the continuous-time Erlang weight |
| $\sigma_w^2$ | Variance of the continuous-time Erlang weight |
| $\eta_w$ | Skew of the continuous-time Erlang weight |
| $K_{\text{CLI}}$ | Number of cascaded (first-order) leaky-integrator filter-components in series, used to generate the Erlang weight |
| $b_m$ | $m$th coefficient of the numerator polynomial of $H(z)$, i.e. $B(z)$ |
| $a_m$ | $m$th coefficient of the denominator polynomial of $H(z)$, i.e. $A(z)$ |
| $K_X$ | Number of monomial terms used in the linear least-squares regressor |
| $K_t$ | Number of temporal derivatives that are output by the linear least-squares regressor |
| $D(e^{i\omega})$ | Frequency response of an ideal (unrealizable) digital low-pass filter |
| $K_\omega$ | Order of frequency-response flatness at dc |
| $k_\omega$ | Order of the derivative of the frequency response with respect to frequency, $k_\omega = 0 \ldots K_\omega$ |
| $\Omega_c$ | Exact cut-off frequency of a continuous-time Butterworth filter (radians per second), determines the -3 dB bandwidth |
| $\omega_c$ | Approximate cut-off frequency of a discrete-time Butterworth filter (radians per sample) |
| $f_c$ | Approximate cut-off frequency of a discrete-time Butterworth filter (cycles per sample) |
| $\widetilde{w}$ | Magnitude of the delayed conjugate product. Is used to form the dynamically weighted average of the phase differences |
| $\|\blacksquare\|$ | Magnitude operator |
| $\angle\blacksquare$ | Angle operator (signal-processing notation) |
| $\arg\{\blacksquare\}$ | Argument operator (same as the angle operator, alternative notation) |
| $\blacksquare^*$ | Complex conjugate |
| $\blacksquare^T$ | Transpose operator |
| $\blacksquare^{-1}$ | Matrix inverse |

## List of acronyms

| | |
|---|---|
| ADC | Analogue-to-Digital Converter |
| BPF | Band-Pass Filter |
| CIC | Cascaded Integrator Comb |
| CLI | Cascaded Leaky Integrator |
| dc | Direct Current, i.e. a frequency of zero Hertz |
| FIR | Finite Impulse Response |
| FFT | Fast Fourier Transform |
| FM | Frequency Modulation |

| HPF | High-Pass Filter |
| --- | --- |
| Hz | Hertz |
| IIR | Infinite Impulse Response |
| LPF | Low-Pass Filter |
| MC | Monte Carlo |
| PSD | Power Spectral Density |
| RMS | Root Mean Squared |
| RMSE | Root Mean Squared Error |
| SNR | Signal-to-Noise Ratio |
| WNG | White-Noise Gain |

List of abbreviations and enumerators

| ALG | Algorithm |
| --- | --- |
| ANG | Angle |
| BUT | Butterworth |
| CPX | Complex |
| DEL | Delay |
| DOM | Domain |
| ERL | Erlang |
| GRP | Group |
| LSQ | Least squares |
| MAG | Magnitude |
| REC | Rectangular |
| WGT | Weight |

# An introduction to instantaneous frequency estimation

The sampled waveform considered in this technical note is modelled as follows:

$x[n] = \psi[n] + \varepsilon_\psi[n]$ where (1)

$x[n]$ is the raw digitized (complex) *waveform* to be processed, a so-called analytic signal, sampled at a uniform rate of $F_s$ (samples per second or Hz) for a sampling period of $T_s = 1/F_s$ (seconds)

$i$ is the imaginary unit

$e$ is Euler's number

$n$ is the sample index

$\varepsilon_\psi[n]$ is complex white noise with real and imaginary parts 'drawn' from a random distribution with a mean of zero and a variance of $\sigma_\varepsilon^2$ and

$\psi[n]$ is a complex *signal*.

The signal is further assumed to have the following form:

$\psi[n] = Ae^{i\theta[n]}$ where (2)

$A$ is the magnitude of the signal, which is assumed to be a (real) constant

$\theta[n]$ is the instantaneous phase or simply the *phase* of the signal. In the classical literature on instantaneous frequency estimation, it is usually assumed that the phase 'progresses' or 'rotates' at a constant rate as

$$\theta[n] = \theta_1 n + \theta_0 \text{ where} \tag{3}$$

$\theta_0$ is the phase offset $\phi_\psi$ of the signal (radians) and

$\theta_1$ is the phase velocity, or phase rate, or the instantaneous frequency, or simply the *frequency* of the signal $\omega_\psi$, i.e. the quantity to be estimated. Note that as an alternative to the *relative angular frequency* $\omega$ (radians per sample), the *relative normalized frequency f* (cycles per sample) is also used here, with $f = \omega/2\pi$. These frequencies are relative to the sampling rate, and they are converted to angular frequency (radians per second) and frequency (cycles per second or Hz) using $F = F_s f$ and $\Omega = F_s \omega$, respectively.

A more general $K_\theta$th-order linear model may also be used to describe the phase progression in both natural and synthetic systems alike (e.g. ultrasonic bat squeaks and pulse compression radars), i.e.

$$\theta[n] = \sum_{k=0}^{K_\theta} \theta_k n^k / k! \,. \tag{4}$$

For a quadratic model with $K_\theta = 2$, the $\theta_2$ parameter is the phase acceleration, or frequency sweep rate, or simply the *chirp* rate (radians per sample per sample). For a cubic model with $K_\theta = 3$, the $\theta_3$ parameter is referred to here as the phase *jerk*.

An obvious way to estimate the $\theta_k$ parameters is to apply the $\arg\{\blacksquare\}$ operator, also known as the angle operator $\angle$ (they are used interchangeably here), to extract raw phase-angle *measurements* from the complex input sequence

$$\angle x[n] = \tilde{\theta}[n] = \theta[n] + \varepsilon_\theta[n] \text{ with} \tag{5a}$$

$$\varepsilon_\theta[n] \sim \varepsilon_\psi[n]/A \tag{5b}$$

for high signal-to-noise ratios, i.e. when $(|\psi|/|\varepsilon_\psi|)^2 \gg 1$ thus $|\varepsilon_\psi/A| \ll 1$ so that $\tan^{-1}\{\varepsilon_\psi/A\} \sim \varepsilon_\psi/A$, where $|\blacksquare|$ is the magnitude operator.

Estimates of the $\theta_k$ parameters in (4), i.e. $\hat{\theta}_k$, are then computed by regressing a $K_\theta$th-order polynomial model to the 'unwrapped' $\tilde{\theta}[n]$ sequence, which is denoted here as $\bar{\theta}[n]$. However, for non-negligible frequencies, the unwrapped measurements grow without bound, which may cause overflows in fixed-point processors and a loss of precision in floating-point processors. A simple solution is to apply the angle operator to delayed conjugate products of the input sequence, i.e. $\arg\{x[n]x^*[n-l]\}$, where $l = 1$ and $\blacksquare^*$ denotes complex conjugation.

When processing the delayed conjugate products *in the absence of noise*, such that $x[n] = \psi[n]$, we have

$$\arg\{\psi[n]\psi^*[n-1]\} = \arg\{A^2 e^{i\theta[n]} e^{-i\theta[n-1]}\} = \arg\{A^2 e^{i\theta[n]-i\theta[n-1]}\}$$

$$= \theta[n] - \theta[n-1] = \bar{\theta}[n] \,. \tag{6}$$

For a constant frequency (i.e. a phase model with $K_\theta = 1$) substitution of (3) into (6) yields

$$\bar{\theta}[n] = \theta[n] - \theta[n-1] = \theta_1 n + \theta_0 - \{\theta_1(n-1) + \theta_0\} = \theta_1 = \omega_\psi \,. \tag{7}$$

In the absence of noise, operating on the delayed conjugate products $\psi[n]\psi^*[n-1]$, eliminates the need for unwrapping when the frequency of the signal is constant ($K_\theta < 2$) because the output of the angle operator is a constant (the frequency to be estimated) and does not exceed $\pm\pi$. However, unwrapping may be required for the $x[n]x^*[n-1]$ sequence when the variance of the noise ($\sigma_\varepsilon^2$) is large and/or when the $|F_\psi|/2$ approaches $F_s/2$ (where $F_\psi$ is the signal frequency), because random perturbations may cause the output of the angle operator to exceed $\pm\pi$. Furthermore, unwrapping is still required (and unbounded time-series handled) when chirped signals ($K_\theta = 2$) are processed if the frequency is swept through $\pm\pi$. The penalty for sequential phase unwrapping is not its computational cost, as it is a very simple logical operation; rather, it is the extra system complexity that follows from the management of potentially unbounded numbers.

Generalizations of this approach that consider a wider range of delays by operating on the discrete-time auto-correlation function, i.e. using $l = 1 \ldots L-1$, have also been proposed. However, they are not considered here because the frequency range of these estimators is drastically reduced from $\pm\pi$ [1] to $\pm\pi/L$ [2] and $\pm 2\pi/(L+1)$ [3], respectively. These methods are claimed to reduce the variance of the frequency estimate at very low SNR. Ways of improving the accuracy of the Kay estimator in similar conditions are described in [9], [10] & [11].

*In the presence of noise*, (1) may be re-written as

$$x[n] = Ae^{i\theta[n]} + \varepsilon_\psi[n] = Ae^{i\theta[n]+i\varepsilon_\theta[n]} \text{ thus} \tag{8a}$$

$$\arg\{x[n]x^*[n-1]\} = \widetilde{\omega}[n] = \omega_\psi + \varepsilon_\omega[n] \text{ where} \tag{8b}$$

$\widetilde{\omega}[n]$ are raw phase-angle-difference measurements.

### General solution

For a signal with a constant frequency ($K_\theta = 1$), a 'smoothed' estimate of the $\theta_1$ parameter, i.e. $\hat{\theta}_1 = \hat{\omega}_\psi$, is now computed by regressing a $K_\theta$th-order polynomial model (in this case, a constant) to the raw sequence $\widetilde{\omega}[n]$ (or the unwrapped sequence $\overline{\omega}[n]$); however, $\varepsilon_\omega[n]$ is no longer white. The regression is realized as a weighted average operation, evaluated over a sliding window, which is usually assumed to be of finite duration, i.e.

$$y[n] = \sum_{m=0}^{M-1} w[m]\widetilde{\omega}[n-m] \text{ where} \tag{9a}$$

$y[n]$ is the output of the smoothing operation at the *current* time of the $n$th sample

$m$ is the delay index (in samples)

$M$ is the window length (in samples) and

$w[m]$ is the relative weight that is applied to each measurement within the sliding window, which is normalized to ensure that $\sum_{m=0}^{M-1} w[m] = 1$.

Note that this averaging operation convolves $\widetilde{\omega}$ (or alternatively, $\overline{\omega}$) with $w$, thus it may be realized via a digital low-pass filter (LPF) with a finite impulse response (FIR) for $M < \infty$ or an infinite impulse response (IIR) for $M = \infty$. The impulse response of this filter $h_{\text{LPF}}[m]$, is equal to the weight applied $w[m]$, and in the IIR case, $w[m] \to 0$ as $m \to \infty$. The frequency response of this filter is $H_{\text{LPF}}(e^{i\omega})$.

For signals with $K_\theta > 1$ (i.e. non-constant frequency) the latency of these smoothing filters should be considered, and their outputs interpreted as follows:

$$y[n] = \hat{\omega}_\psi[n-q] \text{ where} \tag{9b}$$

$\hat{\omega}_\psi[n-q]$ is a *delayed* estimate of the signal's frequency at the time of the $(n-q)$th sample and

$q$ is the passband group delay of the filter (in samples). For low-pass (FIR & IIR) digital filters, it is the negative of the derivative of the phase response, i.e. $\angle H_{\text{LPF}}(e^{i\omega})$, with respect to frequency, evaluated at the dc limit. For linear-phase FIR filters $q = (M-1)/2$.

### Naïve solution

If it could be assumed that $\varepsilon_\omega$ *is white* (it is not) such that its power spectral density (PSD) is constant (i.e. $P_{\varepsilon_\omega}(\omega) = 1/2\pi$), then when $\omega_\psi$ and $A$ are constant, the variance of the frequency-estimate error $(\widehat{\omega}_\psi[n] - \omega_\psi)$ is

$\sigma_{\widehat{\omega}}^2 = v_{\text{LPF}} \sigma_{\widetilde{\omega}}^2$ where (10a)

$\sigma_{\widetilde{\omega}}^2$ is the variance of $\varepsilon_\omega$, with

$\sigma_{\widetilde{\omega}}^2 = \sigma_\varepsilon^2 / A^2$ and (10b)

$v_{\text{LPF}}$ is the white-noise gain (WNG) of the low-pass filter (LPF) or smoother, with

$v_{\text{LPF}} = \sum_{m=0}^{M-1} |w[m]|^2 = \sum_{m=0}^{M-1} |h_{\text{LPF}}[m]|^2$ or from Parseval's theorem (10c)

$v_{\text{LPF}} = \int_{-\pi}^{\pi} |H_{\text{LPF}}(e^{i\omega})|^2 \, d\omega/2\pi$ . (10d)

In this over simplified case, a uniform weight minimizes the variance of the estimator, i.e.

$w[m] = 1/M$ for $m = 0 \ldots M-1$ thus (11a)

$v_{\text{LPF}} = 1/M$ . (11b)

For a sinusoid $\psi$ with a constant magnitude $A$ and a constant frequency $\omega_\psi$ in white noise $\varepsilon_\psi$, this estimate is unbiased; however, for a given $M$ with $M < \infty$, its error variance is not minimized.

### Optimal solution

Kay points out that when $\varepsilon_\psi[n]$ is white, the noise sequence $\varepsilon_\omega[n]$ *is not white*, after the angle operator is applied to the delayed conjugate products $x[n]x^*[n-1]$. Fortunately, the PSD of the coloured noise $P_{\varepsilon_\omega}(\omega)$, is readily derived. During the transformation from delayed conjugate products (a complex signal) to phase differences $\widetilde{\omega}[n]$, the white-noise sequence $\varepsilon_\theta[n]$ passes through a first-order system (a two-point numerical differentiator), which is a digital high-pass filter (HPF) with the following transfer function:

$H_{\text{HPF}}(z) = \frac{z-1}{z} = 1 - 1/z$ thus (12)

$P_{\varepsilon_\omega}(\omega) = |H_{\text{HPF}}(e^{i\omega})|^2 = |1 - e^{-i\omega}|^2$ . (13)

For this *coloured-noise* case, the weight ($w$) that minimizes the variance ($\sigma_{\widehat{\omega}}^2$) is provided in equation (16) of Kay's 1989 paper [1]. In the general case, for an instantaneous-phase model of $K_\theta$th order, it is readily derived using linear regression [4]

$\boldsymbol{w} = \{\boldsymbol{X}^T \boldsymbol{P}_{\varepsilon_\omega}^{-1} \boldsymbol{X}\}^{-1} \boldsymbol{X}^T \boldsymbol{P}_{\varepsilon_\omega}^{-1}$ where (14)

$\boldsymbol{w}$ is an $M \times 1$ vector with elements $w[m]$

$\boldsymbol{X}$ is an $M \times K_\theta$ matrix with regressor vectors as its columns, thus for the zeroth-order regression model (i.e. $K_\theta = 1$), $\boldsymbol{X}$ is a column vector of ones

$\boldsymbol{P}_{\varepsilon_\omega}$ is the (Toeplitz) $M \times M$ covariance matrix of the coloured noise. Elements of the discrete-time coloured-noise autocorrelation function $\boldsymbol{r}_{\varepsilon_\omega}[l]$, run along the $l$th off-diagonal of the covariance matrix, where $l$ is the lag index (in samples), and $\boldsymbol{r}_{\varepsilon_\omega}[l]$ is found via the inverse discrete-time Fourier transform of the noise PSD $P_{\varepsilon_\omega}(\omega)$.

$\blacksquare^T$ is the transpose operator and $\blacksquare^{-1}$ is a matrix inverse.

The error variance of this frequency estimator is found using $v_\text{BPF}$ instead of $v_\text{LPF}$ in (10a) where $v_\text{BPF}$ is the WNG of the band-pass system with the high-pass differentiator and the low-pass smoother in series.

Around its midpoint: the resulting weight is symmetric (for perfect phase linearity) with a quadratic taper, for lower sidelobes (which attenuates high-frequency noise) and a wider bandwidth (for improved tracking of frequencies that are only approximately constant). In ideal conditions (i.e. for a single sinusoid $\psi$ in noise $\varepsilon_\psi$) this tapered weight applies extra attenuation (relative to the rectangular weight) where the PSD $P_{\varepsilon_\omega}$, of the noise sequence $\varepsilon_\omega$, is elevated.

### Pragmatic solution

In non-ideal conditions (i.e. for real hardware, signals and environments) the weight should have extra attenuation where the PSD $P_{\varepsilon_\theta}$, of the noise(+interference) sequence $\varepsilon_\theta$, is elevated. Furthermore, for signal frequencies that are only approximately constant, an averaging filter with a wider bandwidth adjusts to sudden instantaneous-phase discontinuities with greater speed, and tracks 'meandering' or 'sweeping' frequencies with lower bias. Both modifications to the magnitude response of the weight, are achieved by increasing the taper of $w[m]$. Thus, in real environments, parameterizable windows, with a configurable taper, may be required to reduce (steady-state and transient) errors. Furthermore, in real systems, recursive realizations, with a low computational complexity, may be required for online processing at high data-rates.

### Low-pass filter design

The low-pass filters used in the MC simulations of the following section to attenuate the noise $\varepsilon_\omega$, that is added to the phase difference $\omega_\psi = \theta_1 = \theta[n] - \theta[n-1]$ in (8b), are discussed in this section. These digital (FIR & IIR) smoothers are characterized in the time domain by their impulse response $h_\text{LPF}[m] = w[m]$ (see Figure 2), and in the frequency domain by their frequency response $H_\text{LPF}(e^{i\omega})$ (see Figure 3).

The group delay ($q$) and WNG ($v_{LPF}$ & $v_{BPF}$) the various low-pass filters are provided in Table 1 ($v_{LPF}$ is also shown in the legend of Figure 1 & Figure 2). These parameters will be used to interpret the simulation results in the section that follows.

*Table 1. Nominal pass-band group delay (GRP DEL, i.e. q, in samples) and white-noise gain (WNG LPF, i.e. $v_{LPF}$) of the various low-pass smoothing filters. The WNG for the high-pass derivative filter and the low-pass smoothing filter in series is also shown (WNG BPF, i.e. $v_{BPF}$).*

|         | GRP DEL | WNG LPF  | WNG BPF  |
|---------|---------|----------|----------|
| LPF_REC | 12.000  | 0.040000 | 0.003200 |
| LPF_KAY | 12.000  | 0.046291 | 0.000684 |
| LPF_CIC | 12.000  | 0.061457 | 0.001389 |
| LPF_ERL | 14.063  | 0.040000 | 0.000600 |
| LPF_LSQ | 12.956  | 0.049999 | 0.000754 |
| LPF_BUT | 10.397  | 0.081565 | 0.002083 |

https://arxiv.org/abs/2307.00452

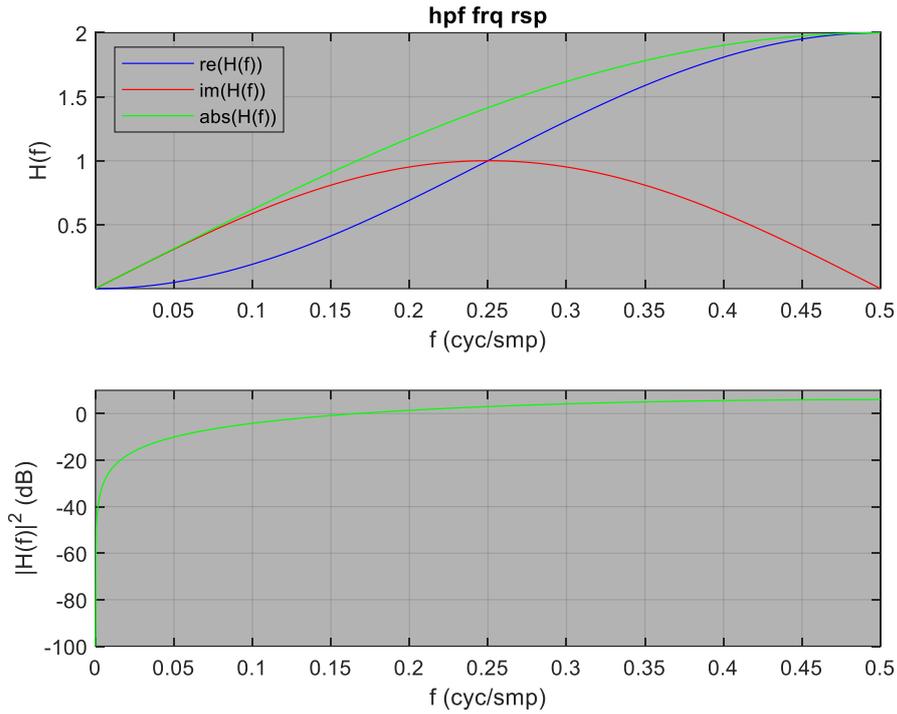

*Figure 1. Frequency response of first-order differentiator – a high-pass filter derived using a two-point difference formula.*

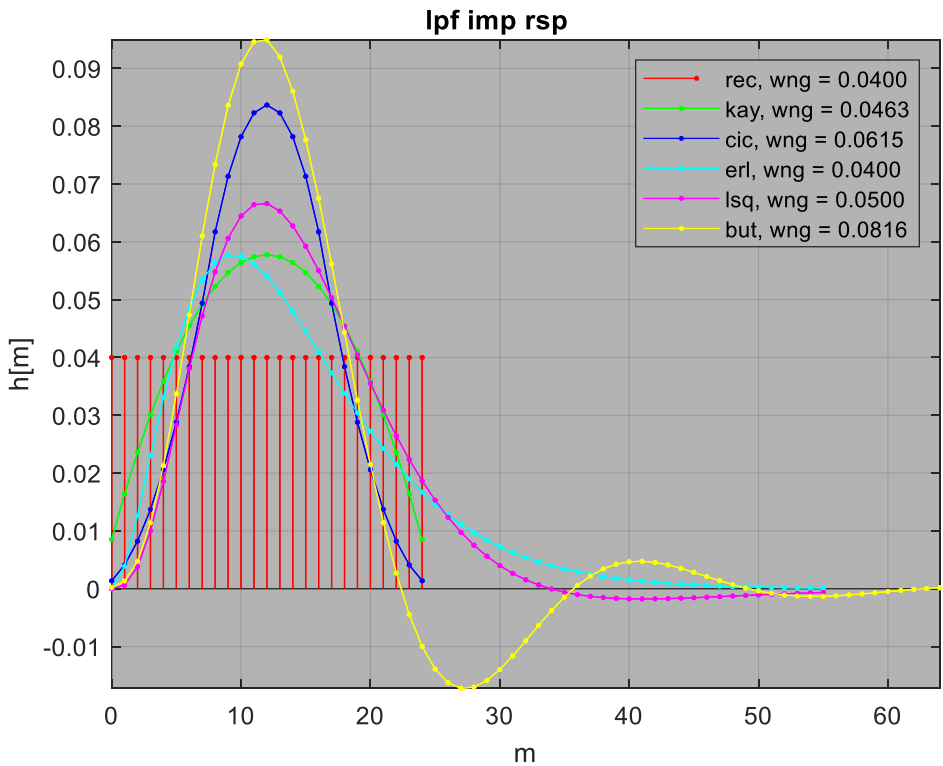

*Figure 2. Impulse responses of low-pass filters. The tails of the IIR filters (cyan, magenta, and yellow, lines) have been truncated.*

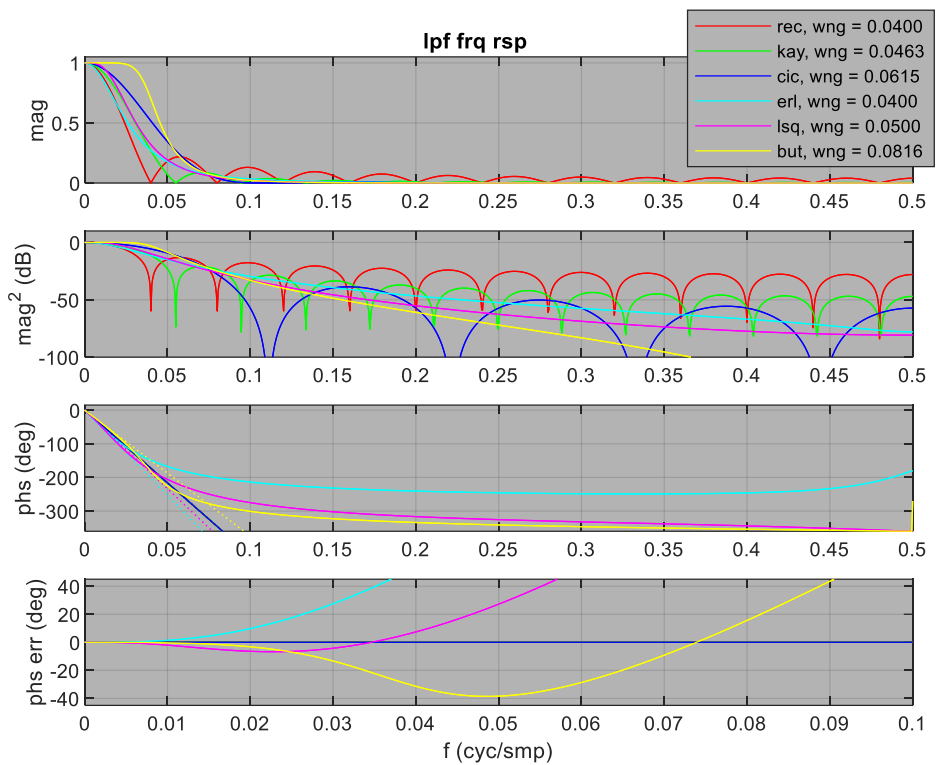

*Figure 3. Frequency responses of low-pass filters. From top to bottom: magnitude (linear scale), squared magnitude (dB scale), phase (in degrees), and phase error, i.e. the deviation from perfect phase linearity (the dotted lines in the phase response subplot). The low-frequency range is shown in the lowermost subplot; the full-frequency range is shown in the other subplots (cycles per sample).*

## Finite Impulse Response (FIR) filters

### Non-recursive

The frequency response of the differentiator in (12), as defined in (13), is shown in Figure 1, i.e. $H_{\text{HPF}}(e^{i\omega})$. For white noise that passes through this system and a signal with a constant frequency, the optimal (Kay) weight for a finite window (with $M = 25$) is shown in Figure 2 (green line). The sub-optimal uniformly weighted (i.e. 'rectangular') window is also shown (red line). These low-pass FIR filters are denoted here using the **LPF_KAY** & **LPF_REC** enumerators.

### Recursive

The taper of the optimal FIR weight may be approximated using an FIR Cascaded Integrator-Comb (CIC) filter [5]. This weight (blue line in Figure 2) is recursively generated by three CIC filter components in series, each with a rectangular impulse response that is nine samples long, i.e. $K_{\text{CIC}} = 3$, $M_{\text{CIC}} = 9$, for $M = K_{\text{CIC}}(M_{\text{CIC}} - 1) + 1 = 25$. The finite impulse response of the CIC filter becomes more Gaussian-like (and more computationally expensive) as $K_{\text{CIC}}$ increases.

## Infinite Impulse Response (IIR) filters

### Recursive

The taper of the optimal FIR weight may also be approximated using an IIR Cascaded Leaky-Integrator (CLI) filter [6]. This weight (cyan line in Figure 2) is recursively generated by forming a linear combination of the outputs of three (first-order) CLI filter components in series, each with an impulse response that decays as an exponential using a real pole at $z = 0.8079$ so it has the same WNG as the rectangular window. The infinite impulse response of the CLI filter becomes more Gaussian-like (and more computationally expensive) as $K_{\text{CLI}}$ increases.

The CLI filter convolves its input with a weight that is a sampled Erlang distribution [6]:

$$h_{\text{LPF}}[m] = w[m] = c_w m^\kappa p^m \text{ where} \tag{15}$$

$\kappa$ is an integer 'shape' parameter that is greater than or equal to zero

$p = e^{-1/\lambda}$ is the 'smoothing' parameter and

$\lambda = -1/\log p$ is the 'timescale' parameter (in samples) and

$c_w$ is a normalizing factor which ensures that $\sum_{m=0}^{\infty} w[m] = 1$.

The mean, variance, and skew, of the continuous-time Erlang distribution, are:

$$\mu_w = (\kappa + 1)\lambda \tag{16a}$$

$$\sigma_w^2 = (\kappa + 1)\lambda^2 \text{ and} \tag{16b}$$

$$\eta_w = 2/\sqrt{\kappa + 1}. \tag{16c}$$

A CLI filter component, with a single pole at $z = p$, generates an Erlang weight with $\kappa = 0$ (a simple exponential decay). For this first-order IIR system

$$c_w = 1 - p \text{ and} \tag{17a}$$

$$H(z) = \frac{b_0 z^1 + b_1 z^0}{a_0 z^1 + a_1 z^0} = \frac{b_0 z^0 + b_1 z^{-1}}{a_0 z^0 + a_1 z^{-1}} \text{ with} \tag{17b}$$

$$b_0 = c_w, \ b_1 = 0 \tag{17c}$$

$$a_0 = 1, \ a_1 = -p. \tag{17d}$$

A third-order CLI filter, with $K_{\text{CLI}}$ poles at $z = p$, generates an Erlang weight with $\kappa = K_{\text{CLI}} - 1$. For this $K_{\text{CLI}}$th-order IIR system

$$c_w = \frac{(1+p)p}{(1-p)^3} \text{ and} \tag{18a}$$

$$H(z) = \frac{b_0 z^3 + b_1 z^2 + b_2 z^1 + b_3 z^0}{a_0 z^3 + a_1 z^2 + a_2 z^1 + a_3 z^0} = \frac{b_0 z^0 + b_1 z^{-1} + b_2 z^{-2} + b_3 z^{-3}}{a_0 z^0 + a_1 z^{-1} + a_2 z^{-2} + a_3 z^{-3}} \text{ with} \tag{18b}$$

$$b_0 = 0, \ b_1 = (1-p)^3/(1+p), \ b_2 = p(1-p)^3/(1+p), \ b_3 = 0 \tag{18c}$$

$$a_0 = 1, \ a_1 = -3p, \ a_2 = 3p^2, \ b_3 = -p^3. \tag{18d}$$

The smoothing parameter ($p$) in (18) was set so that the WNG of this IIR filter is matched to the WNG of the rectangular FIR filter by deriving an expression for the IIR WNG, equating it to (11b), solving for $p$, then using the (real) solution that is on the [0,1] interval.

The impulse response of this filter (see Figure 2, cyan line) has a very gradual taper on the right (as $m \to \infty$) and a smooth rise from zero on the left (as $m \to 0$), which eliminates sidelobes and decreases the high-frequency gain (as $|\omega| \to \pi$); however its asymmetry yields a low-frequency phase response that is not perfectly linear (although its linearity increases as $|\omega| \to 0$, where the gain also approaches unity). As indicated in (16c), the asymmetry (i.e. skew) of this weight decreases as $\kappa$ increases [6].

The CIC filter (with an FIR) and CLI filter (with an IIR) are denoted here using the **LPF_CIC** & **LPF_ERL** enumerators, respectively. Both filters recursively generate the weights for phase-difference averaging, for a fixed computational complexity that is independent of the impulse response duration, which determines the time frame of the average (i.e. the number of samples that are considered), thus the white noise attenuation (as quantified by the WNG). The bandwidth of both filters is determined by the number of cascaded components in series. For a given bandwidth, the WNG of the CIC filter is reduced (in quantized steps) by increasing $M_{\text{CIC}}$; whereas the WNG of the CLI filter is reduced (continuously) by increasing $\lambda$, such that ($p \to 1$).

The FIR and IIR filters described above are all low-pass systems, that attenuate high-frequency noise. Their frequency responses in Figure 3 show that as the taper applied to the weight increases, to make the finite window truncation less severe, the high-frequency gain decreases (i.e. the side-lobes are lowered) and the bandwidth increases (i.e. the main-lobe is broadened).

### Wide-band recursive

It was surmised that other recursive low-pass IIR filters, with a wider bandwidth to better handle deviations from the assumed (constant-frequency) model, may also be used to smooth the phase differences in (8). Such filters, with unity gain at dc (i.e. $\omega = 0$) pass a constant input without attenuation at steady state; however their impulse responses are not necessarily positive for all $m$, therefore their outputs cannot strictly be interpreted as weighted averages. Nevertheless, two candidates were considered, and their properties were investigated in the simulations performed here.

The first (**LPF_LSQ**) regresses a quadratic polynomial model ($K_\theta = 2$) to the phase-difference inputs $\widetilde{\omega}[n]$ using an Erlang weight (also with $\kappa = 2$ and $p = 0.8079$). The IIR filter coefficients were derived using the procedure described in [6], with $K_X = 3$ and $K_t = 1$ for a fifth-order filter.

The second (**LPF_BUT**) was derived by discretizing an analogue Butterworth filter of fourth order and a cut-off frequency of $\Omega_c = 2\pi/M$ (with $M = 25$) so that its $-3$ dB frequency (i.e. its half-power bandwidth) is approximately equal to the $-\infty$ dB frequency (i.e. the first-null bandwidth) of the rectangular window. The discretization was performed using the bilinear transformation (without pre-warping) with $F_s = 1.0$.

The responses of these alternative low-pass IIR filters are also shown in Figure 2 and Figure 3 using magenta and yellow lines, respectively. Figure 3 shows that their bandwidth is wider, and their high-frequency attenuation greater, than the IIR filter derived using three leaky integrators in series. Note that the narrow transition (from the stop band to the pass band) for the fourth-order Butterworth filter (see Figure 3) yields an 'ringing' impulse response (see Figure 2), which is a consequence of the Gibbs' effect; whereas the broader transition band of the fifth-order regression filter yields an impulse response with improved damping.

### Monte-Carlo simulations

Monte Carlo (MC) simulations were performed to investigate various frequency estimation algorithms, with various low-pass smoothing filters. Many different scenarios (i.e. 'runs') were considered and seven of these are used here to illustrate a different aspect of the either the problem, the solution, or the analysis. Each scenario (with one thousand waveform samples, i.e. $N = 1000$) was randomly instantiated one hundred times. Frequency estimation errors were accumulated and the root-mean-squared (RMS) frequency estimation error was computed.

All low-pass filters were initialized to minimize the impact of startup transients. For a given (FIR or IIR) filter, this is done by determining its internal states at steady state (i.e. at the $n \to \infty$ limit) for a unit step input, which is computed analytically (and exactly) using the final-value theorem or numerically (and approximately) in a loop over $n$ until changes to the internal state vector are less than a specified tolerance. The internal state of the filter is then initialized by multiplying the state-vector at infinite time, by the first input sample. However, this does not eliminate start-up transients completely, thus only the samples from $n = (N-1)/8 \ldots (N-1)$ were considered in the RMS Error (RMSE) calculation.

The latency of the filters was considered when the RMSE was computed. This was done by applying a delay to the truth, so it is aligned with the delayed output of the estimator. The high-pass FIR filter stage (with two coefficients) has a group delay of half a sample ($q = 0.5$); the group delays of the low-pass FIR & IIR filter stages are shown in Table 1.

In the first two scenarios (**RUN1**-**RUN2**), the white-noise gain of the filters is used to quantify the RMSE. In the next two scenarios (**RUN3**-**RUN4**), the need for phase unwrapping is illustrated. In the final three scenarios (**RUN5**-**RUN7**), signals that do not match the assumed constant-frequency model are considered, in the absence of additive complex noise, and the frequency response is used to account (qualitatively and quantitatively) for the RMSE. These MC simulations and analysis indicate that the bandwidth of the low-pass filter (i.e. the smoother) is a critical parameter and that in practice, it should be set to reach the desired balance between random errors and bias errors. On the one hand, a wider bandwidth amplifies the white noise that falls within the passband of the low-pass filter, thus random errors are increased. On the other hand, it also improves the tracking of non-ideal signals that are encountered in practice, which reduces bias errors.

### RUN1: Low noise

In this scenario, complex noise $\varepsilon_\psi$ with $\sigma_\varepsilon = 0.01$ was added to a complex signal with $A = 1$, a constant frequency of $\theta_1 = \omega_\psi = 2\pi f_\psi$, with $f_\psi = 0.1$ and a randomly generated phase offset $\theta_0$. A random instantiation of this scenario is shown in Figure 4. The sampled waveform $x[n]$, is in the upper subplot (real part in blue, imaginary part in red, and magnitude in green); the real and imaginary parts of the delayed conjugate products $x[n]x^*[n-1]$, are in the lower subplot, along with the noise-free products $\psi[n]\psi^*[n-1]$ (black lines). In Kay's alternative approach, the lowpass filters are applied in the complex domain (to the delayed conjugate products) instead of the angle domain, thus the smoothed outputs of the various low-pass filters described in the previous section are also plotted here (coloured lines, see Figure 3 for legend). The angle domain is depicted in Figure 5 with true frequency $f_\psi$ (black line), raw angle measurements $\tilde{f}[n] = \tilde{\omega}[n]/2\pi$ (black dots) and frequency estimates $\hat{f}_\psi[n] = \hat{\omega}_\psi[n]/2\pi$ (coloured lines). The six low-pass filters (**LPF_REC**, **LPF_KAY**, **LPF_CIC**, **LPF_ERL**, **LPF_LSQ**, **LPF_BUT**, line colour) were used to filter the phase differences in the *angle domain* (**DOM_ANG**, dashed lines) and the delayed conjugate products in the *complex domain* (**CPX_DOM**, solid lines), for twelve different frequency estimators, or algorithms (**ALG01** … **ALG12**).

The RMSE of each algorithm is shown in Table 2, along with the expected RMSE (Exp. RMSE) computed using (10), using the WNG values (i.e. $v_{\mathrm{BPF}}$) in Table 1. The observed and expected values are in good agreement: 2-3 significant figures for **DOM_ANG** and 1-2 significant figures for **DOM_CPX**; with all **DOM_CPX** errors slightly higher than the corresponding **DOM_ANG** errors, in agreement with the literature [1].

It is difficult to see the different steady-state error characteristics of each estimator in Figure 5, therefore a magnified plot is provided in Figure 6. Even though **LPF_REC** has the lowest WNG (see $v_{\mathrm{LPF}}$ in Table 1), its elevated sidelobes (see Figure 3) amplify the high-frequency noise of the high-pass differentiator (see Figure 1) thus it has the greatest variance at steady state ($\sigma_{\hat{\omega}}^2$) as predicted (it has the largest $v_{\mathrm{BPF}}$ in Table 1).

https://arxiv.org/abs/2307.00452

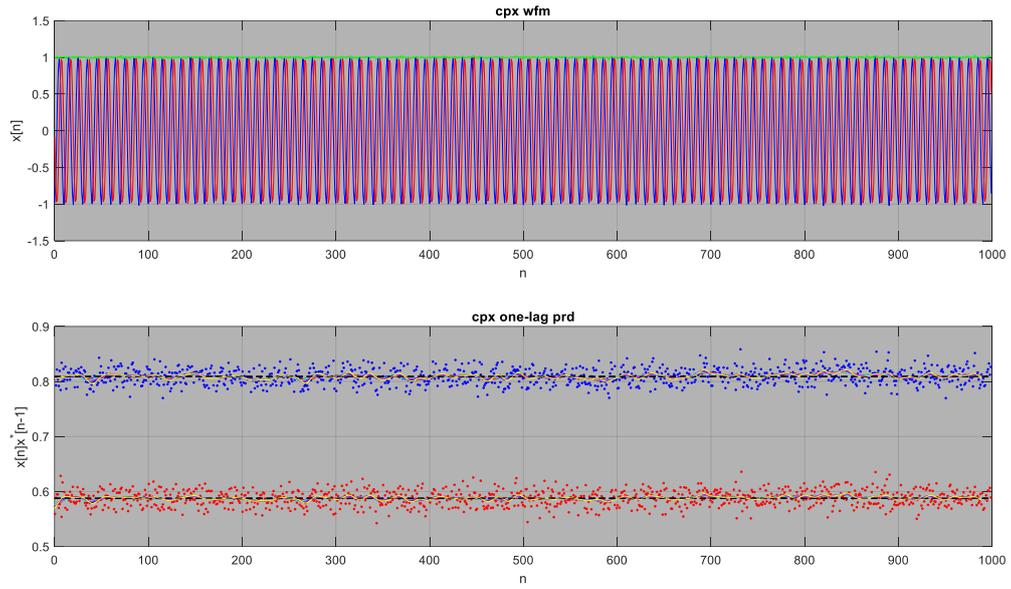

*Figure 4. Raw waveform and product waveform of a RUN1 instantiation.*

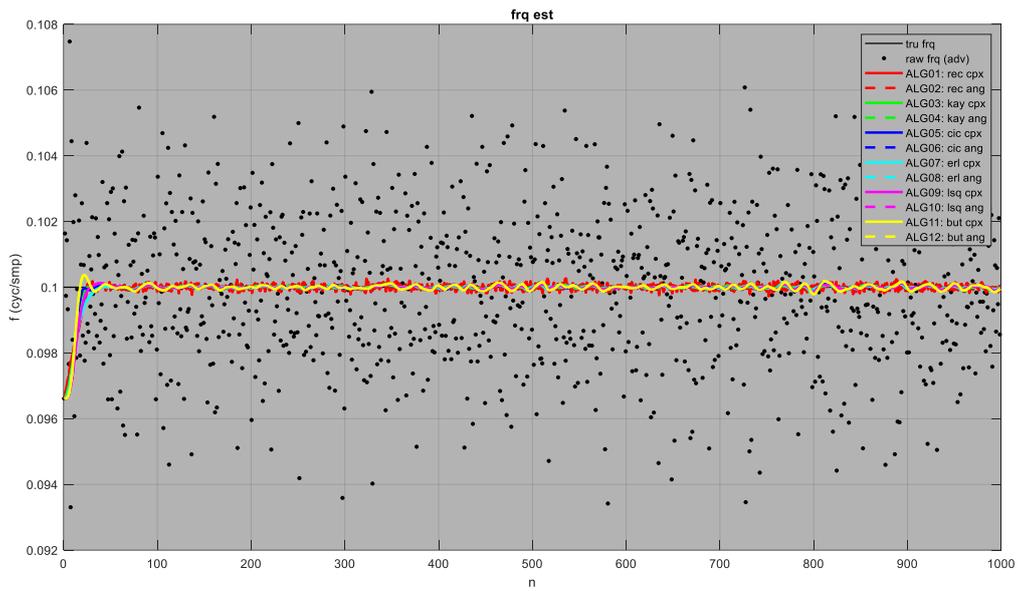

*Figure 5. Angle measurements and frequency estimates for the waveform shown in Figure 4.*

https://arxiv.org/abs/2307.00452

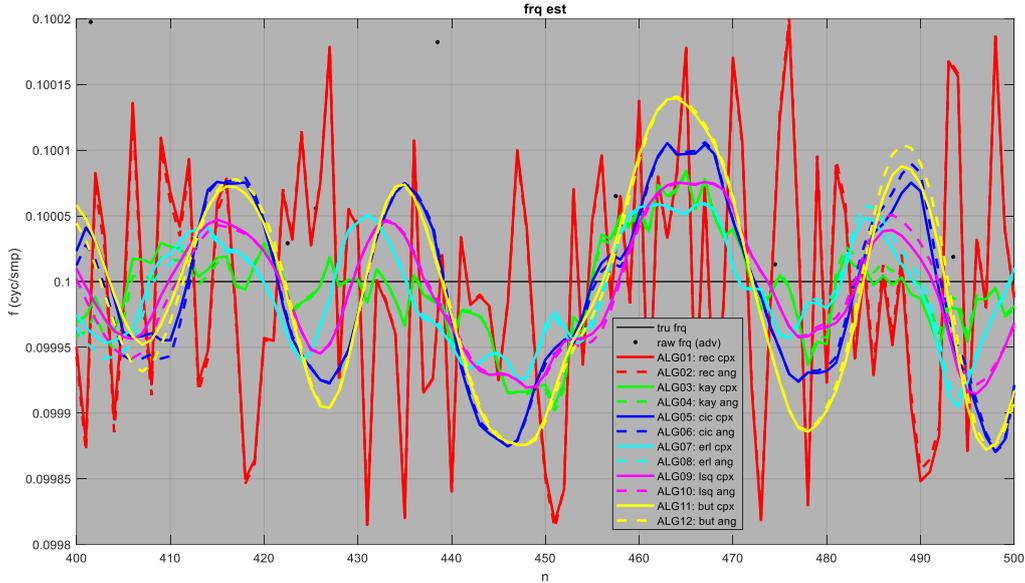

*Figure 6. Frequency estimates in Figure 5 (with true frequency) at high magnification.*

*Table 2. Observed frequency errors (radians per sample) for the RUN1 MC simulations (Sim. RMSE) and the expected errors calculated from the WNG (Exp. RMSE) for various estimators.*

|         | Sim. RMSE for **DOM_CPX** |          | Sim. RMSE for **DOM_ANG** |          | Exp. RMSE |
|---------|---------------------------|----------|---------------------------|----------|-----------|
| **LPF_REC** | 5.672E-04 | (**ALG01**) | 5.666E-04 | (**ALG02**) | 5.657E-04 |
| **LPF_KAY** | 2.656E-04 | (**ALG03**) | 2.638E-04 | (**ALG04**) | 2.615E-04 |
| **LPF_CIC** | 3.783E-04 | (**ALG05**) | 3.765E-04 | (**ALG06**) | 3.727E-04 |
| **LPF_ERL** | 2.490E-04 | (**ALG07**) | 2.474E-04 | (**ALG08**) | 2.450E-04 |
| **LPF_LSQ** | 2.794E-04 | (**ALG09**) | 2.776E-04 | (**ALG10**) | 2.747E-04 |
| **LPF_BUT** | 4.629E-04 | (**ALG11**) | 4.608E-04 | (**ALG12**) | 4.564E-04 |

## RUN2: High noise

This scenario is the same as the previous scenario (**RUN1**); however, the noise ($\sigma_\varepsilon$) is increased from 0.01 to 0.10 (see Figure 7). The predicted errors (Exp. RMSE) and the observed errors (Sim. RMSE) are again in good agreement (see Table 3) for the filters that are applied in the angle domain (**DOM_ANG**); however, the observed errors in the complex domain (**DOM_CPX**) are much worse than predicted, in agreement with the literature [1].

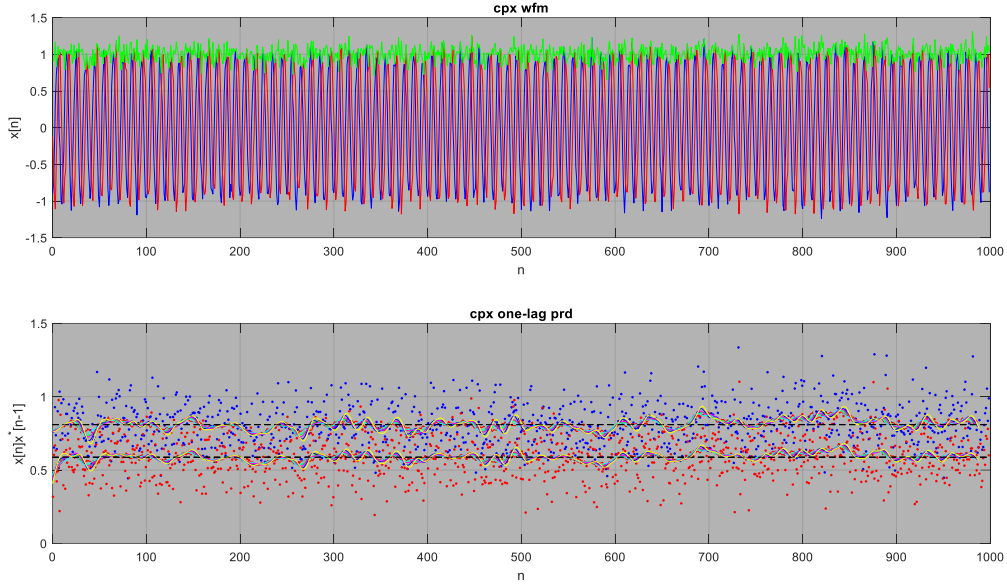

*Figure 7. Raw waveform and product waveform of a RUN2 instantiation.*

*Table 3. Observed frequency errors (radians per sample) for the RUN2 MC simulations (Sim. RMSE) and the expected errors calculated from the WNG (Exp. RMSE) for various estimators.*

|         | Sim. RMSE for **DOM_CPX** |          | Sim. RMSE for **DOM_ANG** |          | Exp. RMSE |
|---------|---------------------------|----------|---------------------------|----------|-----------|
| **LPF_REC** | 6.331E-03 | (**ALG01**) | 5.687E-03 | (**ALG02**) | 5.657E-03 |
| **LPF_KAY** | 4.038E-03 | (**ALG03**) | 2.647E-03 | (**ALG04**) | 2.615E-03 |
| **LPF_CIC** | 5.168E-03 | (**ALG05**) | 3.780E-03 | (**ALG06**) | 3.727E-03 |
| **LPF_ERL** | 3.757E-03 | (**ALG07**) | 2.483E-03 | (**ALG08**) | 2.450E-03 |
| **LPF_LSQ** | 4.213E-03 | (**ALG09**) | 2.786E-03 | (**ALG10**) | 2.747E-03 |
| **LPF_BUT** | 6.185E-03 | (**ALG11**) | 4.626E-03 | (**ALG12**) | 4.564E-03 |

## RUN3: High frequency and high noise

This scenario is a the same as the previous scenario (**RUN1**); however, the noise ($\sigma_\varepsilon$) is doubled (from 0.1 to 0.2) and the frequency of the signal ($f_\psi$) is increased (from 0.1 to 0.4). With these modifications, the phase-difference measurements ($\tilde{f}$) occasionally cross the $f = \pm 1/2$ boundary, as discussed above, in the paragraph that follows (7). The RMSE for all **DOM_CPX** low-pass filters is less than the corresponding **DOM_ANG** RMSE in all cases, except for **LPF_LSQ** (see Table 4). In the **DOM_CPX** case, unwrapping is not required because complex numbers (not angles) are being smoothed. In the **DOM_ANG** case, the angle measurements that wrap around are not being 'interpreted' correctly by the smoother, which causes large 'excursions' in the wrong direction (see Figure 8). The **LPF_LSQ** smoother was applied using a 'hand-coded' linear-state-space recursion (see [6]) which allows unwrapping to be applied on each update using

$$\bar{\omega}[n] = \hat{\omega}_\psi[n-1] + \arg\{e^{i\tilde{\omega}[n]} e^{-i\hat{\omega}_\psi[n-1]}\} \text{ where} \tag{19}$$

$\bar{\omega}[n]$ is the unwrapped angle measurement to be smoothed.

All other smoothers were applied using a call to Matlab's in-built `filter()` function (realized via FFTs in the FIR case), which is does not provide an opportunity to apply this unwrapping logic; however, all smoothers could have been applied using a linear-state-space recursion, if so desired.

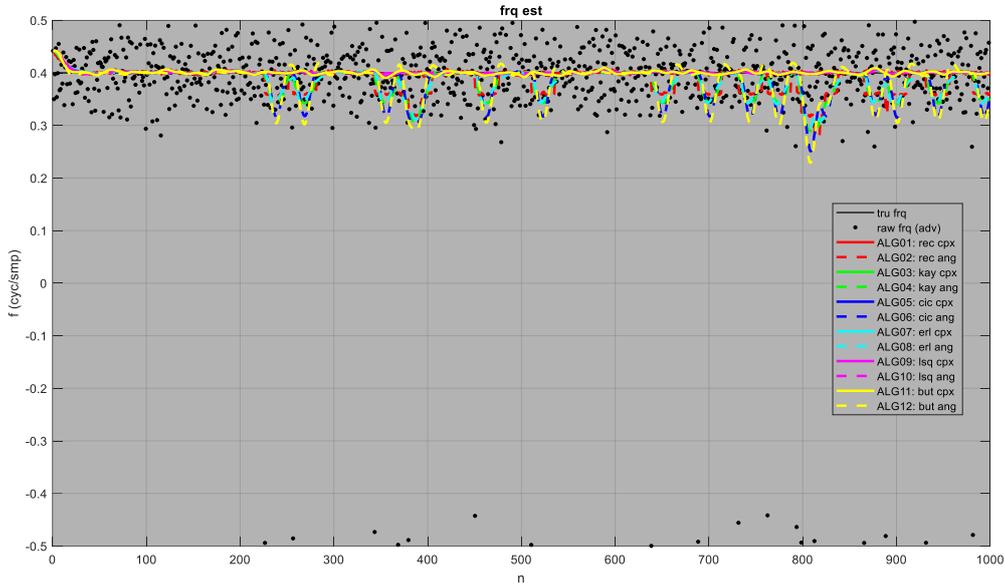

*Figure 8. Angle measurements and frequency estimates for a RUN3 instantiation.*

*Table 4. Observed frequency errors (radians per sample) for the RUN3 MC simulations.*

|  | Sim. RMSE for **DOM_CPX** |  | Sim. RMSE for **DOM_ANG** |  |
|---|---|---|---|---|
| **LPF_REC** | 1.609E-02 | (**ALG01**) | 1.811E-01 | (**ALG02**) |
| **LPF_KAY** | 1.342E-02 | (**ALG03**) | 1.909E-01 | (**ALG04**) |
| **LPF_CIC** | 1.618E-02 | (**ALG05**) | 2.130E-01 | (**ALG06**) |
| **LPF_ERL** | 1.246E-02 | (**ALG07**) | 1.812E-01 | (**ALG08**) |
| **LPF_LSQ** | 1.399E-02 | (**ALG09**) | 5.625E-03 | (**ALG10**) |
| **LPF_BUT** | 1.903E-02 | (**ALG11**) | 2.390E-01 | (**ALG12**) |

## RUN4: High frequency and very high noise

This scenario is a the same as the previous scenario (**RUN3**); however, the noise ($\sigma_\varepsilon$) is again doubled (from 0.2 to 0.4). Wrap-around events are now much more frequent and the incorrectly interpreted angle measurements lead to a large and enduring bias towards zero (see Figure 9 and Table 5).

*Table 5. Observed frequency errors (radians per sample) for the RUN4 MC simulations.*

|  | Sim. RMSE for **DOM_CPX** |  | Sim. RMSE for **DOM_ANG** |  |
|---|---|---|---|---|
| **LPF_REC** | 5.243E-02 | (**ALG01**) | 1.007E+00 | (**ALG02**) |
| **LPF_KAY** | 5.245E-02 | (**ALG03**) | 1.018E+00 | (**ALG04**) |
| **LPF_CIC** | 6.230E-02 | (**ALG05**) | 1.045E+00 | (**ALG06**) |
| **LPF_ERL** | 4.836E-02 | (**ALG07**) | 1.007E+00 | (**ALG08**) |
| **LPF_LSQ** | 5.478E-02 | (**ALG09**) | 3.049E-02 | (**ALG10**) |
| **LPF_BUT** | 7.412E-02 | (**ALG11**) | 1.079E+00 | (**ALG12**) |

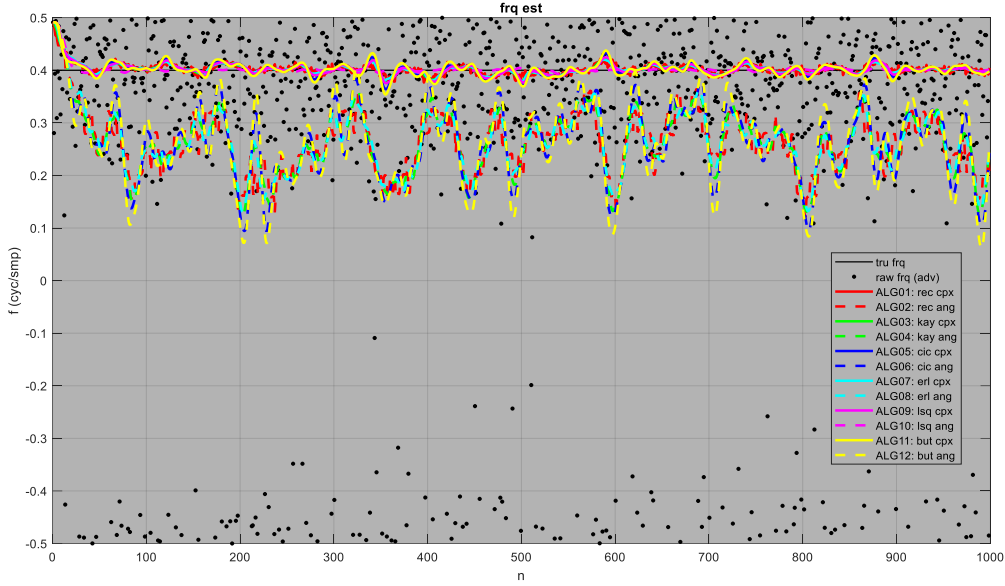

*Figure 9. Angle measurements and frequency estimates for a RUN4 instantiation.*

## RUN5: Linear frequency sweep with no noise

A linear frequency-modulated signal (LFM) or 'up-chirp' is considered in this scenario. The frequency is swept from $f = -0.4$ at $n = 0$ to $f = 0.4$ at $n = N-1$, thus $\theta_1 = -2\pi \times 0.4$ and $\theta_2 = 2\pi \times 0.8/(N-1)$. No noise is added in this scenario ($\sigma_\varepsilon = 0$). The complex waveform is shown in Figure 10; the frequency measurements and estimates in Figure 11; and error metrics in Table 6. The delayed conjugate product is a sinusoid with a frequency of $\theta_2$ (see lower subplot of Figure 10). The smoothing filters are all designed to pass non-oscillating (i.e. constant) signals with no magnitude scaling (i.e. gain or loss). However, for oscillating signals the magnitude scaling is equal to the magnitude response of the low-pass filter at that frequency, i.e. $|H_{LPF}(e^{i\omega})|$ and the phase delay is equal to the phase response of the low-pass filter at that frequency i.e. $\angle H_{LPF}(e^{i\omega})$. When the angle operator is applied to the output of the low-pass filter (applied in the complex domain), magnitude is ignored thus magnitude shifts have no impact on the frequency estimate; however, phase shifts may introduce a bias. For FIR filters with a perfectly linear phase response over all frequencies, the delay experienced is the same for all frequencies. For IIR filters with a nearly linear-phase response in the passband, the delay is not the same for all frequencies. The phase-response error ($\theta_\Delta$) and the expected RMSE was therefore computed using

$$\theta_\Delta = \arg\{H_{LPF}(e^{i\theta_2})e^{iq\theta_2}\} \text{ where} \tag{20a}$$

$\angle H_{LPF}(e^{i\omega})$ is the realized phase response of the low-pass filter and

$\angle e^{-iq\omega}$ is the desired phase response of the low-pass filter then

$$\text{RMSE} = |\theta_\Delta|. \tag{20b}$$

The low-frequency phase-response error of each low-pass filter is shown in the lowermost subplot of Figure 3. The value of these curves at $f = \theta_2/2\pi = 0.8/(N-1) = 8.008 \times 10^{-4}$ is equal to $\theta_\Delta$ (in degrees). The FIR filters (**LPF_REC**, **LPF_KAY** & **LPF_CIC**) all have perfect phase-linearity thus a zero phase-response error. Of the IIR filters (**LPF_ERL**, **LPF_LSQ** & **LPF_BUT**), the Butterworth filter (**LPF_BUT**) has the widest bandwidth thus the lowest error. The RMSE expected from the analysis above (Exp. RMSE) is provided in Table 6 and it agrees with the RMSE observed in the simulations (Sim.

https://arxiv.org/abs/2307.00452https://arxiv.org/abs/2307.00452

RMSE), for the smoothers applied in the complex domain (**DOM_CPX**), to 3-4 significant figures. For the smoothers applied in the angle domain (**DOM_ANG**), the input is a polynomial, not a sinusoid, therefore this analysis is not applicable there; however, an alternative approach is presented in the scenario that follows.

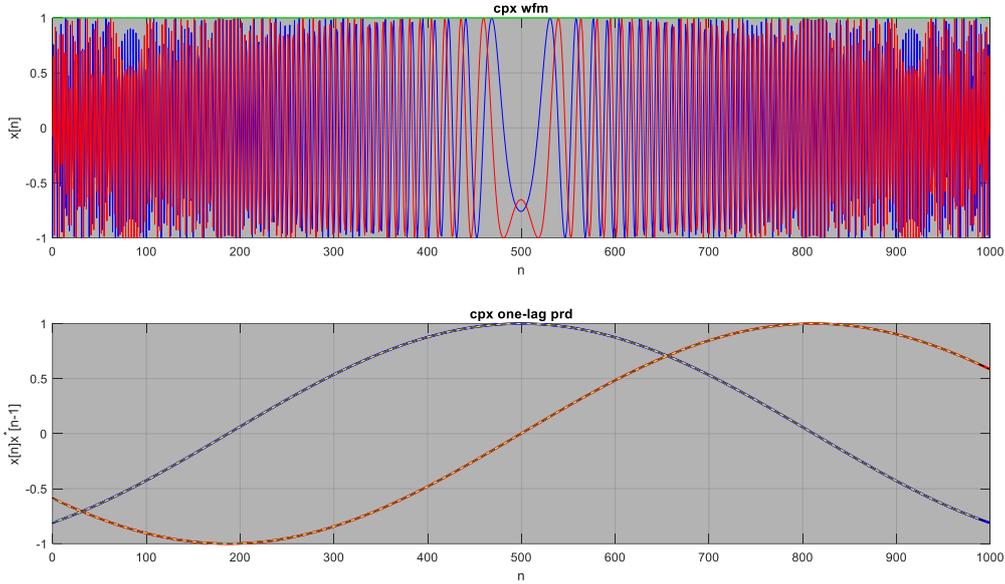

*Figure 10. Raw waveform and product waveform for a RUN5 instantiation.*

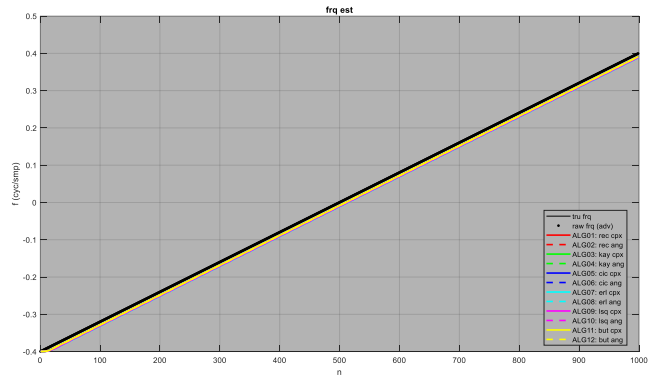

*Figure 11. Angle measurements and frequency estimates for the waveform shown in Figure 10.*

*Table 6. Observed frequency errors (radians per sample) for the RUN5 MC simulations (Sim. RMSE) and the expected errors calculated from the phase response (Exp. RMSE) for various estimators.*

|  | Sim. RMSE for **DOM_CPX** | | Sim. RMSE for **DOM_ANG** | | Exp. RMSE |
|---|---|---|---|---|---|
| **LPF_REC** | 5.381E-15 | (**ALG01**) | 5.398E-15 | (**ALG02**) | 0.000E+00 |
| **LPF_KAY** | 1.698E-15 | (**ALG03**) | 1.701E-15 | (**ALG04**) | 0.000E+00 |
| **LPF_CIC** | 2.360E-15 | (**ALG05**) | 2.382E-15 | (**ALG06**) | 0.000E+00 |
| **LPF_ERL** | 1.312E-05 | (**ALG07**) | 1.837E-12 | (**ALG08**) | 1.312E-05 |
| **LPF_LSQ** | 2.696E-05 | (**ALG09**) | 1.199E-10 | (**ALG10**) | 2.696E-05 |
| **LPF_BUT** | 3.007E-06 | (**ALG11**) | 1.023E-08 | (**ALG12**) | 3.006E-06 |

## RUN6: Quadratic frequency sweep with no noise

A quadratic frequency sweep is considered in this scenario. The frequency is swept from $f = 0.0$ at $n = 0$, up to $f = 0.25$ at $n = (N-1)/2$, then down to $f = -0.25$ at $n = N-1$ thus $\theta_1 = 0.0$, $\theta_2 = 2\pi \times 1.2513 \times 10^{-3}$ and $\theta_3 = -2\pi \times 1.5030 \times 10^{-6}$. No noise is added in this scenario ($\sigma_\varepsilon = 0$). The complex waveform is shown Figure 12; the frequency measurements and estimates in Figure 13; and error metrics in Table 7. For this non-linear frequency sweep, the delayed complex product is *not* a sinusoid (see lower subplot of Figure 12); however, the angle measurements *are* a low-order (quadratic) polynomial (see Figure 13). The bandwidth of the low-pass filters accounts for the RMSE observed in the simulations, i.e. the wider the bandwidth, the greater the accuracy of the instantaneous frequency estimate (in the absence of noise). For smoothers applied in the complex domain, the magnitude response is used to quantify the filter bandwidth (see Table 8); whereas the flatness of the frequency response at dc is used in the angle domain (see Table 9).

In the former case (**DOM_CPX**), the magnitude response $\left|H_{LPF}(e^{i\omega})\right|$ is evaluated at $\omega = 2\pi \times 1/4 \times 1/M$, i.e. at one-quarter of the first-null bandwidth of the rectangular weight (**LPF_REC**). The closer this realized gain is to the 'ideal' value of unity, the lower the expected RMSE.

In the latter case (**DOM_ANG**), the dc flatness is evaluated as follows [6],[7],[8]:

For a 'perfectly flat' low-pass filter, with unity gain in the passband, and a group delay of $q$, the desired derivatives of the frequency response, with respect to frequency, evaluated at dc, are

$$\left\{\frac{d^{k_\omega}}{d\omega^{k_\omega}}D(e^{i\omega})\right\}\bigg|_{\omega=0} = (-iq)^{k_\omega} \text{ for } k_\omega = 0 \ldots \infty \,. \tag{21}$$

This ideal or desired response $D(e^{i\omega})$, is unrealizable for a digital filter; however, low-pass filters with $K_\omega$th-order flatness are realizable, i.e.

$$\left\{\frac{d^{k_\omega}}{d\omega^{k_\omega}}H_{LPF}(e^{i\omega})\right\}\bigg|_{\omega=0} = (-iq)^{k_\omega} \text{ for } k_\omega = 0 \ldots K_\omega \tag{22}$$

and (in the absence of noise or interference) such filters track a polynomial of $K_\omega$th degree with zero error at steady state (i.e. as $n \to \infty$).

In this scenario, the smoothers applied in the angle domain must track a quadratic frequency sweep, i.e. a polynomial of second degree, therefore a low-pass filter with second-order flatness ($K_\omega = 2$) is required. The "Realized" and "Ideal" $k_\omega = 2$ derivatives at dc for the various low-pass filters are provided in Table 9, along with their difference (i.e. the "Error"). For $k_\omega = 0$ and $k_\omega = 1$ all filters have the desired dc derivatives, thus they are not shown in the table. The smaller the difference, the lower the RMSE. Only the **LPF_LSQ** and **LPF_BUT** filters have the required flatness for the perfect tracking of a quadratic signal at steady state.

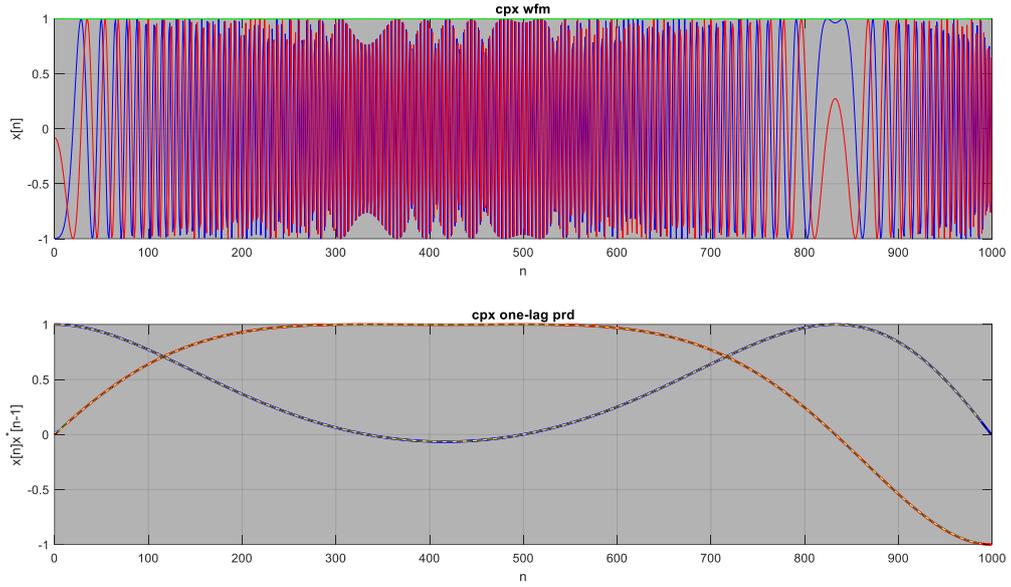

*Figure 12. Raw waveform and product waveform for a RUN6 instantiation.*

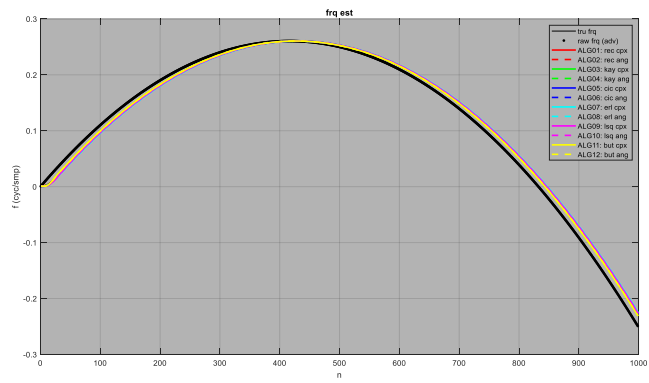

*Figure 13. Angle measurements and frequency estimates for the waveform shown in Figure 12.*

*Table 7. Observed frequency errors (radians per sample) for the RUN6 MC simulations (Sim. RMSE) for various estimators.*

|         | Sim. RMSE for **DOM_CPX** |           | Sim. RMSE for **DOM_ANG** |           |
|---------|---------------------------|-----------|---------------------------|-----------|
| **LPF_REC** | 4.916E-04 | (**ALG01**) | 4.919E-04 | (**ALG02**) |
| **LPF_KAY** | 3.179E-04 | (**ALG03**) | 3.181E-04 | (**ALG04**) |
| **LPF_CIC** | 1.896E-04 | (**ALG05**) | 1.897E-04 | (**ALG06**) |
| **LPF_ERL** | 6.408E-04 | (**ALG07**) | 6.233E-04 | (**ALG08**) |
| **LPF_LSQ** | 6.999E-05 | (**ALG09**) | 7.870E-07 | (**ALG10**) |
| **LPF_BUT** | 8.807E-06 | (**ALG11**) | 7.882E-07 | (**ALG12**) |

Table 8. Observed frequency errors (radians per sample) for the RUN6 MC simulations (Sim. RMSE) and the magnitude response at f=1/(4M), realized and Ideal, for various low-pass filters.

|         | Sim. RMSE for **DOM_CPX** |          | Realized  | Ideal     | Error      |
|---------|---------------------------|----------|-----------|-----------|------------|
| **LPF_REC** | 4.916E-04             | (**ALG01**) | 9.005E-01 | 1.000E+00 | -9.954E-02 |
| **LPF_KAY** | 3.179E-04             | (**ALG03**) | 9.352E-01 | 1.000E+00 | -6.478E-02 |
| **LPF_CIC** | 1.896E-04             | (**ALG05**) | 9.612E-01 | 1.000E+00 | -3.881E-02 |
| **LPF_ERL** | 6.408E-04             | (**ALG07**) | 8.827E-01 | 1.000E+00 | -1.173E-01 |
| **LPF_LSQ** | 6.999E-05             | (**ALG09**) | 9.726E-01 | 1.000E+00 | -2.745E-02 |
| **LPF_BUT** | 8.807E-06             | (**ALG11**) | 1.000E+00 | 1.000E+00 | -7.649E-06 |

Table 9. Observed frequency errors (radians per sample) for the RUN6 MC simulations (Sim. RMSE) and the second derivative of the frequency response evaluated at dc for the ideal (desired) and realized low-pass filters.

|         | Sim. RMSE for **DOM_ANG** |          | Realized   | Ideal      | Error       |
|---------|---------------------------|----------|------------|------------|-------------|
| **LPF_REC** | 4.919E-04             | (**ALG02**) | -1.960E+02 | -1.440E+02 | -5.200E+01  |
| **LPF_KAY** | 3.181E-04             | (**ALG04**) | -1.776E+02 | -1.440E+02 | -3.360E+01  |
| **LPF_CIC** | 1.897E-04             | (**ALG06**) | -1.640E+02 | -1.440E+02 | -2.000E+01  |
| **LPF_ERL** | 6.233E-04             | (**ALG08**) | -2.637E+02 | -1.978E+02 | -6.592E+01  |
| **LPF_LSQ** | 7.870E-07             | (**ALG10**) | -1.679E+02 | -1.679E+02 | -1.866E-09  |
| **LPF_BUT** | 7.882E-07             | (**ALG12**) | -1.081E+02 | -1.081E+02 | -1.321E-08  |

## RUN7: Modulated signal with no noise

A phase-and-magnitude modulated sinusoid of constant frequency of $f_\psi = 0.2$, and no additive noise, is considered in this scenario i.e.

$x[n] = \chi[n]\psi[n]$ where (23)

$\psi[n] = e^{i\theta[n]}$ with $K_\theta = 1$ and $\theta_1 = 2\pi \times 0.2$ and

$\chi[n]$ is the modulator, which was synthesized by low-pass filtering complex noise. The real and imaginary parts of the noise were pseudo-randomly generated according to a Gaussian distribution with a mean of zero and a variance of one. The complex noise was passed through a fourth-order Butterworth filter with a cut-off frequency of $\Omega_c = 2\pi/100$.

The waveforms generated in two MC-simulation instantiations are shown in Figure 14 & Figure 17.

An alternative form of frequency estimator (**ANG_DOM+MAG_WGT**, dash-dot lines) is also considered for these modulated waveforms, whereby the raw phase-differences are averaged using an additional 'dynamic' weight $\widetilde{w}$, that is equal to the relative magnitudes of the delayed conjugate products. Thus (9) is modified as follows:

$y[n] = \sum_{m=0}^{M-1} w[m]\widetilde{w}[n-m]\widetilde{\omega}[n-m] / \sum_{m=0}^{M-1} w[m]\widetilde{w}[n-m]$ where (24)

$\widetilde{w}[n] = |x[n]x^*[n-1]|$ .

The RMSE of all algorithms is provided in Table 10.

Smoothing in the complex domain (**DOM_CPX**) yields a lower error than smoothing in the angle domain (**ANG_DOM**), except for when the wider bandwidth IIR filters are used (**LPF_LSQ** & **LPF_BUT**). These IIR filters perform poorly because their wider bandwidths and narrow transition bands yield impulse responses with ringing tails that cross zero (see Figure 2), thus their impulse responses cannot be interpreted as weights. This feature did not lead to significant performance degradation in the previous scenarios, but it does in this scenario whenever the magnitude of the waveform is relatively small (see Figure 15 & Figure 18). The RMSE trend in Table 10 is also visible in the magnified plot of an error spike in Figure 16. The error spike near $n = 160$ in Figure 18, for the Butterworth filter applied in the complex domain (**ALG_11**), is particularly bad. The smoothed outputs for all filters applied in the complex domain is shown in Figure 19. In this plot it is apparent that the spurious impulse response of the Butterworth filter used in **ALG_11** yields erroneous averages, where the signal magnitude is relatively small, when the smoothed real and imaginary outputs cross zero and each other.

The RMSE of the magnitude-weighted algorithms (**ANG_DOM+MAG_WGT**) is very similar to the complex domain algorithms (**CPX_DOM**); although for all low-pass filters, the RMSE is slightly worse.

For an ideal signal, i.e. constant magnitude and constant frequency, smoothing in the angle domain (**ANG_DOM**) yields a lower variance than smoothing in the complex domain (**CPX_DOM**), because magnitude is disregarded in the angle domain. This effectively constrains the estimator and removes a degree of freedom, which reduces the impact of random noise. However, in this scenario with a variable magnitude, this assumption no longer holds, thus the relative advantage becomes a disadvantage.

Table 10. Observed frequency errors (radians per sample) for the RUN7 MC simulations (Sim. RMSE) and the expected errors calculated from the phase response (Exp. RMSE) for various estimators.

|  | Sim. RMSE for **DOM_CPX** | | Sim. RMSE for **DOM_ANG** | | Sim. RMSE for **DOM_ANG+MAG_WGT** | |
|---|---|---|---|---|---|---|
| **LPF_REC** | 3.212E-02 | (**ALG01**) | 5.040E-02 | (**ALG02**) | 3.213E-02 | (**ALG13**) |
| **LPF_KAY** | 3.514E-02 | (**ALG03**) | 5.330E-02 | (**ALG04**) | 3.515E-02 | (**ALG14**) |
| **LPF_CIC** | 3.889E-02 | (**ALG05**) | 5.708E-02 | (**ALG06**) | 3.891E-02 | (**ALG15**) |
| **LPF_ERL** | 3.125E-02 | (**ALG07**) | 5.052E-02 | (**ALG08**) | 3.126E-02 | (**ALG16**) |
| **LPF_LSQ** | 1.328E-01 | (**ALG09**) | 5.464E-02 | (**ALG10**) | | |
| **LPF_BUT** | 1.756E-01 | (**ALG11**) | 6.228E-02 | (**ALG12**) | | |

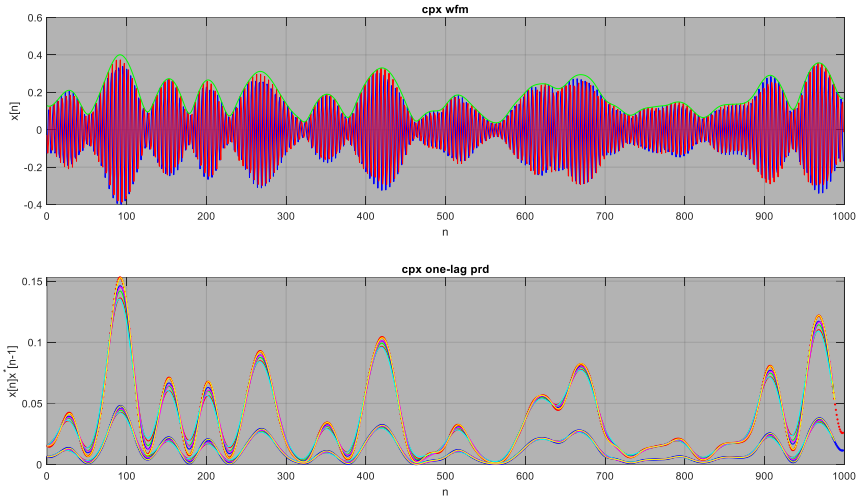

Figure 14. Raw waveform and product waveform for the first RUN7 instantiation.

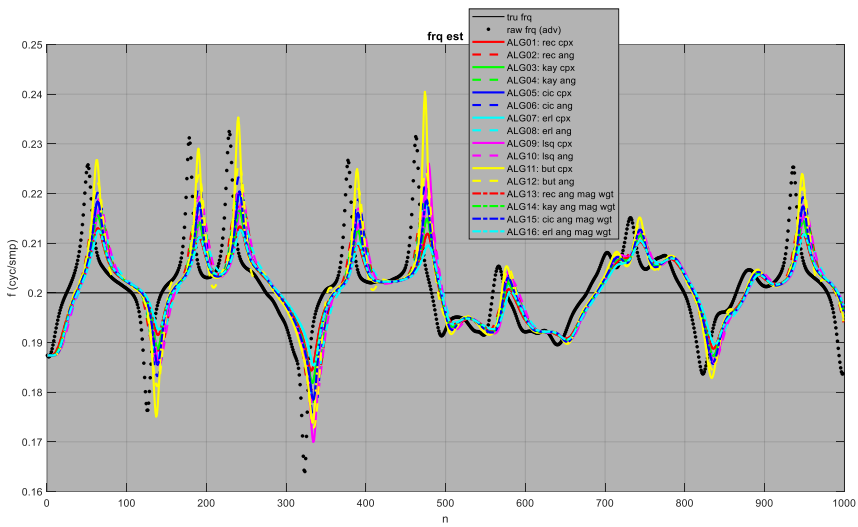

Figure 15. Angle measurements, true frequency, and frequency estimates for the waveform shown in Figure 14.

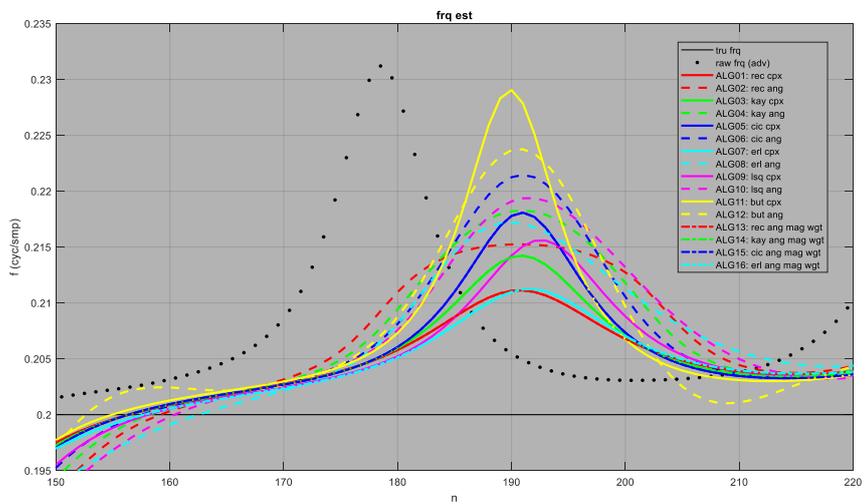

Figure 16. Frequency estimates in Figure 15 (with true frequency and angle measurements) at high magnification.

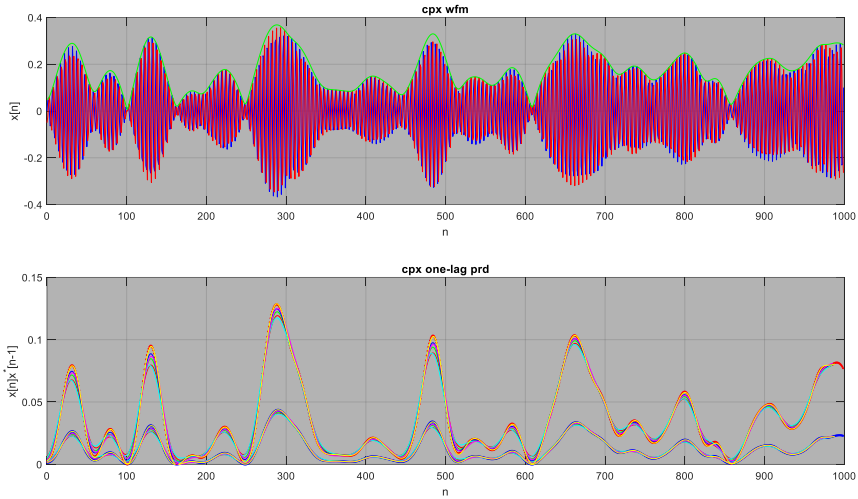

Figure 17. Raw waveform and product waveform for the second RUN5 instantiation.

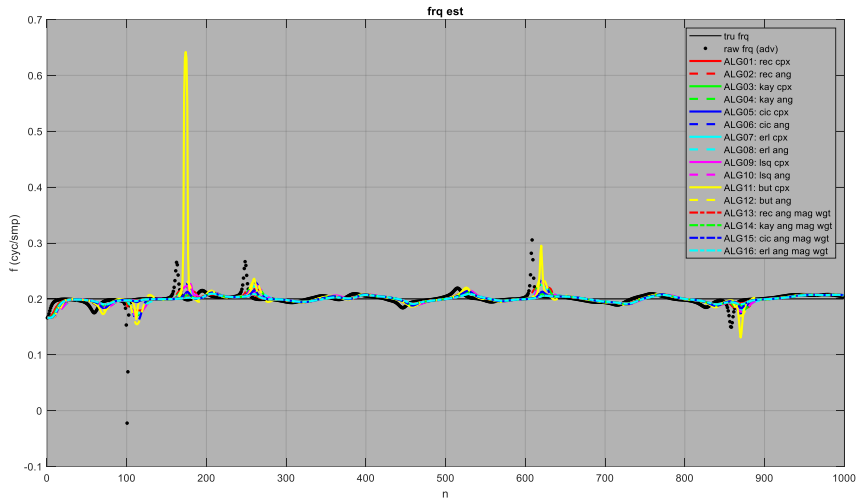

Figure 18. Angle measurements, true frequency, and frequency estimates for the waveform shown in Figure 17.

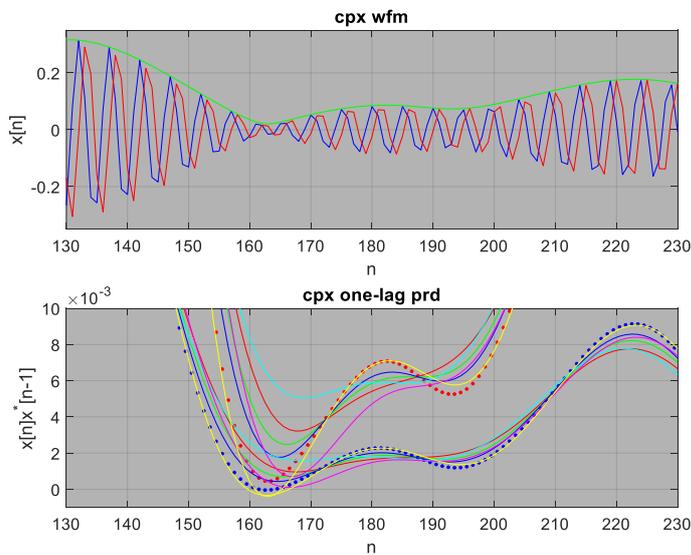

Figure 19. Raw waveform (upper subplot). Product waveform (lower subplot) with the outputs of various low-pass filters (applied in the complex domain). A detail of Figure 18 at high magnification. The error for the frequency estimate of ALG11 spikes around n = 162 where the (real and imaginary) outputs of the LPF_BUT filter (yellow lines) cross over zero.

## Conclusion

The following main points were considered, illustrated, and discussed, in this technical note:

- For a sinusoid in white noise, the error variance of the frequency estimator may be derived (quantitatively) from the high-pass frequency response of the phase differentiating filter (a two-point difference) and the low-pass frequency response of the weight that is used to compute their sliding average (a smoother).
- When averaging phase differences, computed from delayed conjugate products, unwrapping is required in the following (albeit unlikely) 'corner' cases:
    - The (white) noise variance is very high
    - The (constant) signal frequency is near the Nyquist rate
    - The signal frequency is not constant and passes the Nyquist rate.
- In theory, Kay's tapered weight minimizes the error variance for an average that is computed over a finite window (applied via an FIR smoother); however, in low SNR conditions, long windows are required, which may be computationally prohibitive at very high data rates.
- FIR Cascaded Integrated Comb (CIC) filters and IIR Cascaded Leaky Integrator (CLI) filters are reasonable approximations of Kay's optimal weight, with very low computational complexity due to their recursive realization. These filters may indeed be optimal in practice when the power spectral density of the complex noise is elevated at higher frequencies and/or when high-frequency interference is present (due to their lower high-frequency gain), and when the frequency and magnitude of the signal is not constant (due to their wider bandwidth).
- For signals with a linearly swept frequency, in the absence of additive noise, the error variance of a frequency estimator may be predicted (quantitatively) from the phase response of the low-pass filter (when it is applied in the complex domain).
- For other signals of non-constant frequency, in the absence of additive noise, the error variance of a frequency estimator may be inferred (qualitatively) from the magnitude response of the smoother (when it is applied in the complex domain) and from the flatness of the response at the dc limit (when it is applied in the angle domain).
- At high SNR, the error variance of frequency estimators that involve the smoothing (or averaging) in the complex domain (i.e. before the angle operator is applied) is only slightly worse than those that do smoothing in the angle domain, but much worse at low SNR. However, the former class of estimator does not require phase unwrapping while smoothing and it is more robust when processing non-ideal signals of variable instantaneous phase-rate (i.e. frequency) and magnitude, e.g. for (phase and/or magnitude) modulated signals or swept-frequency signals.
- When smoothing is done in the angle domain, the magnitude weighting of phase differences, yields a frequency estimation error variance that is only slightly greater than when it is done in the complex domain.